\documentclass[twocolumn]{aastex62}
\usepackage{CJK}
\listofchanges

\begin{document}
\begin{CJK*}{UTF8}{gbsn}

\title{Evidence of AGN activity in the gamma-ray emission from two starburst galaxies}

\correspondingauthor{Fang-Kun Peng}
\email{pengfangkun@163.com}
\correspondingauthor{Xiang-Yu Wang}
\email{xywang@nju.edu.cn}
\correspondingauthor{Qi-Jun Zhi}
\email{qjzhi@gznu.edu.cn}

\author[0000-0001-7171-5132]{Fang-Kun Peng (彭方坤)}
\affiliation{Department of Physics, Anhui Normal University, Wuhu, Anhui 241000, PR China; } 
\affiliation{School of Astronomy and Space Science, Nanjing University, Nanjing 210023, China}
\affiliation{Guizhou Provincial Key Laboratory of Radio Astronomy and Data Processing, Guizhou Normal University, Guiyang 550001, China}

\author{Hai-Ming Zhang}
\affiliation{School of Astronomy and Space Science, Nanjing University, Nanjing 210023, China}
\affiliation{Key laboratory of Modern Astronomy and Astrophysics (Nanjing University), Ministry of Education, Nanjing 210023, China}

\author{Xiang-Yu Wang}
\affiliation{School of Astronomy and Space Science, Nanjing University, Nanjing 210023, China}
\affiliation{Key laboratory of Modern Astronomy and Astrophysics (Nanjing University), Ministry of Education, Nanjing 210023, China}

\author{Jun-Feng Wang}
\affiliation{Department of Astronomy and Institute of Theoretical Physics and Astrophysics, Xiamen University, Xiamen, Fujian 361005, China}

\author{Qi-Jun Zhi}
\affiliation{Guizhou Provincial Key Laboratory of Radio Astronomy and Data Processing, Guizhou Normal University, Guiyang 550001, China} 

\begin{abstract}
Starburst galaxies are huge reservoirs of cosmic rays (CRs) and these CRs convert a significant fraction of their energy into gamma-rays by colliding with the  interstellar medium (ISM).
The produced GeV gamma-ray emission is temporally stable and the flux is found to correlate well with indicators of star formation rates, such as the total infrared (IR) luminosity $L_{\rm IR}$ and monochromatic radio continuum luminosity at 1.4 GHz $L_{\rm 1.4\ GHz}$, i.e., following $L_{\gamma}-L_{\rm IR}$ and $L_{\gamma}-L_{\rm 1.4\ GHz}$ relations.
Recently, gamma-ray excesses are reported to be spatially coincident with  two starburst galaxies NGC 3424 and UGC 11041  in the fourth \textsl{Fermi} Large Area Telescope (LAT) source catalog (4FGL).
Different from other starburst galaxies detected by \textsl{Fermi}-LAT, we find that the gamma-ray emission associated with NGC 3424 and UGC 11041 show significant flux variability.
With relatively weak infrared and radio emission,  NGC 3424 and UGC 11041 appear as outliers of the $L_{\gamma}-L_{\rm IR}$ and $L_{\gamma}-L_{\rm 1.4\ GHz}$ relations of starburst galaxies.
These results suggest that NGC 3424 and UGC 11041 may harbor obscured active galactic nuclei (AGNs) and the AGN activities provide the dominant contribution to the gamma-ray emission as compared to that provided by the starburst activities.
\end{abstract}

\keywords{cosmic rays -- gamma-rays: ISM -- galaxies: star formation--galaxies: Seyfert}

\section{Introduction}
Nearby star-forming and starburst galaxies have been identified to be GeV-TeV gamma-ray sources \citep{2009Sci...326.1080A,2009Natur.462..770V,2010ApJ...709L.152A,2010A&A...523A..46A,2010A&A...512A...7A,2010A&A...523L...2A,2012ApJ...755..164A,2016A&A...586A..71A,2014ApJ...794...26T,2016ApJ...821L..20P,2016ApJ...823L..17G,2017ApJ...836..208A}.
Cosmic rays (CRs) accelerated by supernova remnants (SNRs) or stellar winds inevitably interact with the interstellar medium (ISM) and produce neutral pions (schematically written as $p+p\rightarrow \pi^0$+other products), which subsequently in turn decay into high-energy gamma-rays ($\pi^0\rightarrow \gamma+\gamma$).
The gamma-ray emission produced in this process is expected to be stable, as has been demonstrated in some starburst galaxies (e.g., \citealp{2012ApJ...755..164A}).
Meanwhile, in the connections that link the star formation and CRs in starburst galaxies, some authors have proposed scaling relationships between star formation rates (SFRs) and gamma-ray luminosities ($L_{\gamma}$) \citep{2002ApJ...575L...5P,2004ApJ...617..966T,2007ApJ...654..219T,2007APh....26..398S,2010MNRAS.403.1569P,2011ApJ...734..107L}.
SFR indicators include the total infrared (IR) luminosity $L_{\rm IR}$ at $8-1000 \ \micron $ converted to a SFR based on dust-processed stellar light \citep{1998ApJ...498..541K}, and radio continuum  luminosity at $1.4$ GHz produced by synchrotron-emitting CR electrons \citep{2001ApJ...554..803Y}.
With the accumulation of \textsl{Fermi}-Large Area Telescope (LAT) data, the correlations between gamma-ray luminosities and SFR indicators are first found in \citet{2010A&A...523L...2A}, confirmed by later studies with larger sample \citep{2012ApJ...755..164A}, and extended to a larger luminosity scale with the detection of gamma-ray emissions from a luminous infrared galaxy NGC 2146 \citep{2014ApJ...794...26T} and an ultra-luminous infrared galaxy Arp 220 \citep{2016ApJ...821L..20P,2016ApJ...823L..17G}.

Among the star-forming and starburst galaxies observed by \textsl{Fermi}-LAT, NGC 1068 and NGC 4945 deserve special attention as these two galaxies show both circumnuclear starbursts activities and radio-quiet AGNs with low-level jets \citep{2010A&A...524A..72L,2012ApJ...755..164A}.
AGNs and star-forming (or starburst) galaxies, the leading components of the extragalactic source population in the GeV range detected by \textsl{Fermi}-LAT \citep{2012ApJ...755..164A,2015ApJS..218...23A,2015ApJ...810...14A,2016ApJ...826..190A,2017MNRAS.466..952A,2019arXiv190210045T}, produce gamma-ray emission by different physical processes.
The former objects arise from intermittent accretion onto the central massive black holes, while the latter objects generate diffuse gamma-ray emission by interaction between CRs and ISM.
Observationally, none of these starburst galaxies show evidence of gamma-ray variability, while variability is commonly seen in  \textsl{Fermi}-LAT AGNs.
The gamma-ray spectra of NGC 1068 and NGC 4945 are similar to those of two typical starburtst galaxies of M82 and NGC 253.
What's more, NGC 1068 and NGC 4945 obey the $L_{\gamma}-L_{\rm IR}$ and $L_{\gamma}-L_{\rm 1.4\ GHz}$ relations for starburst galaxies \citep{2012ApJ...755..164A}.
In this respect, the gamma-ray emissions in NGC 1068 and NGC 4945 are probably produced dominantly by CRs interaction (hence starburst activity) \citep{2012ApJ...755..164A}, although the AGN-driven outflows in NGC 1068 and NGC 4945 might provide  complementary contribution to the gamma-ray emission under some model assumptions (e.g., \citealp{2010A&A...524A..72L,2016A&A...596A..68L}).
Another controversial source is Circinus galaxy.
\citet{2013ApJ...779..131H} reported the discovery of gamma-ray excess from the Circinus galaxy using the early four years of \textsl{Fermi}-LAT observations, and found that this observed gamma-ray luminosity exceeds the luminosity expected from CRs interactions in the ISM and inverse Compton radiation from the radio lobes. But considering its location near the Galactic plane with $b = -3\arcdeg .8$, systematic error related to uncertainties in the model for the Galactic diffuse and point sources emission should be evaluated with new data \citep{2019arXiv190210045T,2019arXiv190504723G}.

\citet{2019arXiv190210045T} recently released the latest catalog  based on the first eight years of \textsl{Fermi}-LAT science data. Relative to the previous 3FGL catalog \citep{2015ApJS..218...23A}, the 4FGL benefits from many improvements, including twice longer exposure, Pass 8 data, new model of diffuse Galactic emission, and updates on likelihood analysis and association procedure.
The new 4FGL catalog contains two new sources in spatial association with two starburst galaxies, NGC 3424 and UGC 11041, listed in the IRAS catalog\citep{2003AJ....126.1607S}.
NGC 3424 and UGC 11041 were ranked No. 430 and No. 186 out of 629 IRAS sources in the distribution of $L_{\rm IR}$ values \citep{2003AJ....126.1607S}, respectively.
The IR emission of these two objects were so  weak that we do not expect they will be bright GeV emitters involving star formation.
The radio flux density at 1.4 GHz of NGC 3424 is $\sim 161$ mJy \citep{1992ApJS...79..331W}.
Such a source would be unusually weak by comparison with that of \textsl{Fermi}-LAT star-forming and starburst galaxies.
In this paper, in order to investigate the physical origin of the GeV excesses, we  analyze the \textsl{Fermi}-LAT data of NGC 3424 and UGC 11041, performing a detailed analysis of the temporal and spectral characteristics, and examining whether they obey the relationships between gamma-ray luminosity and SFR tracers, like $L_{\gamma}-L_{\rm IR}$ and $L_{\gamma}-L_{\rm 1.4\ GHz}$ relations for starburst galaxies.

The rest of this paper is structured as follows. In Section 2, we describe the data reduction and results of the \textsl{Fermi}-LAT observations. In section 3, we study the correlations involving gamma-ray luminosities of starburst galaxies. We give our discussions in Section 4 and conclusions in Section 5.

\section{\textsl{Fermi} data reduction}
\subsection{Data selection}
We downloaded the updated \textsl{Fermi}-LAT Pass 8 SOURCE data towards to sources of interest, i.e., NGC 3424 and UGC 11041, for a period of about 10.5 years (MET 239557417 - MET 573163231).
A binned maximum likelihood analysis was performed on a region of interest (ROI) with a radius $10\arcdeg$ centered on the "R.A." and "decl." of each source.
Recommended event type for data analysis was "FRONT+BACK" (evtype=3).
We applied a maximum zenith angle cut of $z_{\rm zmax} = 90\arcdeg$ to reduce the effect of the Earth albedo background.
The standard gtmktime filter selection with expression of ($\rm DATA\_QUAL > 0 \ \&\& \ LAT\_CONFIG == 1)$ was set.
A source model was generated containing the position and spectral definition for all the point
sources and diffuse emission from the 4FGL \citep{2019arXiv190210045T} within $15\arcdeg$ of the ROI center.
The Galactic and extragalactic diffuse models were $\rm gll\_iem\_v07.fits$ and $\rm iso\_P8R3\_SOURCE\_V2\_v1.txt$, respectively. The energy dispersion correction was made when events energies extended down to $100$ MeV were took in to consideration.

In order to get a convergence of fit and good error estimate on parameters in the time resolved and energy resolved analysis, we handled the model file in two different ways. One was to fix the spectral shape of all source, and the other was to fix the spectral parameters for the sources with $>= 7\arcdeg$ degrees away from ROI center.
The bright gamma-ray sources, Galactic diffuse and isotropic emission components were left free in the fitting procedure.
It was found that the two methods did not make any statistical difference to the results.

\subsection{Spatial analysis}
First, we optimized a point source location using the likelihood test-statistic (TS) around the NGC 3424 region, and checked the results assessed in \textsl{Fermi}-LAT 4FGL \citep{2019arXiv190210045T}.
We performed an unbinned method for photons with energies $> 200$ MeV.
We estimated the best-fit gamma-ray emission position with the tool \textsl{gtfindsrc}: $(162\arcdeg .922, 32\arcdeg .8767)\pm 0\arcdeg .04$.
We found the final guess position was in accordance well with that in 4FGL in the error box.
Then, the test-statistic map around the NGC 3424 region was made. Based on the test-statistic map, there was little residual gamma-ray emission nearby, indicating that one point source in the model could depict the gamma-ray emission well in the vicinity of NGC 3424 region.

Since AGNs dominate the population of extragalactic sources detected by \textsl{Fermi}-LAT \citep{2015ApJ...810...14A,2019arXiv190210045T},  we checked possible alternative candidates for this point source in the following main AGN catalogs: the CRATES Flat-Spectrum Radio Source Catalog \citep{2007ApJS..171...61H}, the Veron Catalog of Quasars AGN, the 13th Edition \citep{2010A&A...518A..10V}, the Candidate Gamma-Ray Blazar Survey Source Catalog \citep{2008ApJS..175...97H}, and Roma-BZCAT Multi-Frequency Catalog of Blazars \citep{2009A&A...495..691M}.
No AGN candidate is found within the positional uncertainty of the point source.
Only an IRAS point source associated with NGC 3424 is located within the region.
With only a separation of $0\arcdeg .03$ between the gamma-ray point source and NGC 3424, it is reasonable to ascribe the high-energy emission from the gamma-ray point source to NGC 3424.
Following the similar steps, we examined the data of UGC 11041.
The derived power law index and flux of NGC 3424 and UGC 11041 are compatible with results presented in \textsl{Fermi}-LAT 4FGL.
Although no  AGN candidate is found, a flat-spectrum radio source MG2 J175448+3442 is very close to UGC 11041 and could be an alternative candidate associated with the gamma-ray excess \citep{2019arXiv190210045T}.
\textsl{Fermi}-LAT 4FGL lists UGC 11041 as a variable source in the \textsl{Fermi}-LAT band, which is confirmed by our following data analysis procedure, so the gamma-ray emission most likely arises from an AGN \citep{2019arXiv190210045T}.

\subsection{Light curve and spectral analysis}
We generated a set of time bins for the light curve of photons with energy $>200$ MeV.
In the first trial, the full observation period was divided linearly into five equal time bins.
Each time bin was fitted by a separate maximum likelihood analysis.
The results are shown in Table \ref{tablelc5}.
One could see that the gamma-ray emission in the third time bin was stronger than the other bins by a factor of about one order of magnitude for these two sources, indicating  a clear variability.
\textsl{Fermi}-LAT has worse angular resolution at low energy.
In order to reduce contamination from the nearby sources, we increased the energies of events to above $400$ MeV and $800$ MeV.
The gamma-ray flux variations of NGC 3424 and UGC 11041 are still detected.
The unbinned method and Front photons are used to perform the light curve analysis to check the results, and we find that the trend of gamma-ray emission clustering in the third time bin is not changed.

\begin{table*}[!htp]
\centering
\begin{tabular}{cccccccc}
\hline
\hline
Name&Energies & & Aug 4, 2008  & Sep 15, 2010  & Oct 27, 2012 & Dec 8, 2014 & Jan 18, 2017 \\
    &         & & Sep 15, 2010 & Oct 27, 2012  & Dec 8, 2014  & Jan 18, 2017 & Mar 1, 2019 \\
\hline
NGC 3424&0.2-100 GeV &TS &2 & 0 & 56 & 9 & 0 \\
&            &Flux &$1.46\pm 1.53 $ & 6.49 & $2.28\pm 0.46$ & $5.93\pm 0.47$ &3.69  \\
&           & &$\times10^{-10}$ & $\times10^{-10}$ & $\times10^{-9}$ & $\times10^{-10}$ & $\times10^{-10}$ \\
\hline
&0.4-100 GeV &TS & 4 & 0 & 49 & 7 & 0 \\
&            &Flux &$1.41\pm 1.26 $ & 1.82 & $1.29\pm 0.04$ & $2.98\pm 0.16$ &1.75 \\
&            & &$\times10^{-10}$ & $\times10^{-10}$ & $\times10^{-9}$ & $\times10^{-10}$ & $\times10^{-10}$ \\
\hline
&0.8-500 GeV & TS & 3 & 0 & 49 & 10 & 0 \\
&            &Flux &$4.76\pm 3.81 $ & 1.72 & $5.57\pm 1.28$ & $1.66\pm 0.75$ &8.54 \\
&           & &$\times10^{-11}$ & $\times10^{-10}$ & $\times10^{-10}$ & $\times10^{-10}$ & $\times10^{-11}$ \\
\hline
UGC 11041 & 0.2-100 GeV&TS &0 & 6 & 97 & 7 & 2 \\
&            &Flux &$4.88 $ & $1.35\pm 0.63$ & $7.41\pm 0.88$ & $2.78\pm 1.07$ &$8.33\pm 6.02$  \\
&          & &$\times10^{-10}$ & $\times10^{-9}$ & $\times10^{-9}$ & $\times10^{-9}$ & $\times10^{-10}$ \\
\hline
&0.4-100 GeV &TS & 0 & 5 & 73 & 4 & 2 \\
&            &Flux & 3.72 & $5.35\pm 2.67$ & $2.51\pm 0.35$ & $5.20\pm 2.55$ &$3.02\pm 2.30$ \\
&           & &$\times10^{-10}$ & $\times10^{-10}$ & $\times10^{-9}$ & $\times10^{-10}$ & $\times10^{-10}$ \\
\hline
&0.8-500 GeV & TS & 0 & 6 & 36 & 3 & 3 \\
&            &Flux & 2.13  & $2.70\pm 1.28$ & $6.72\pm 1.42$ & $2.00\pm 1.21$ & $1.93\pm 1.13$ \\
&            & &$\times10^{-10}$ & $\times10^{-10}$ & $\times10^{-10}$ & $\times10^{-10}$ & $\times10^{-10}$ \\
\hline
\hline
\end{tabular}
\caption{Light curve of NGC 3424 and UGC 11041 with five time bins. Flux is in unit of $\rm ph \ cm^{-2} \ s^{-1}$. All error bars represent $1 \sigma$ uncertainties. The data points with $\rm TS < 1$ are given upper limits at the 95\% confidence level.}
\label{tablelc5}
\end{table*}

\begin{figure*}
\centering
\includegraphics[scale=0.45]{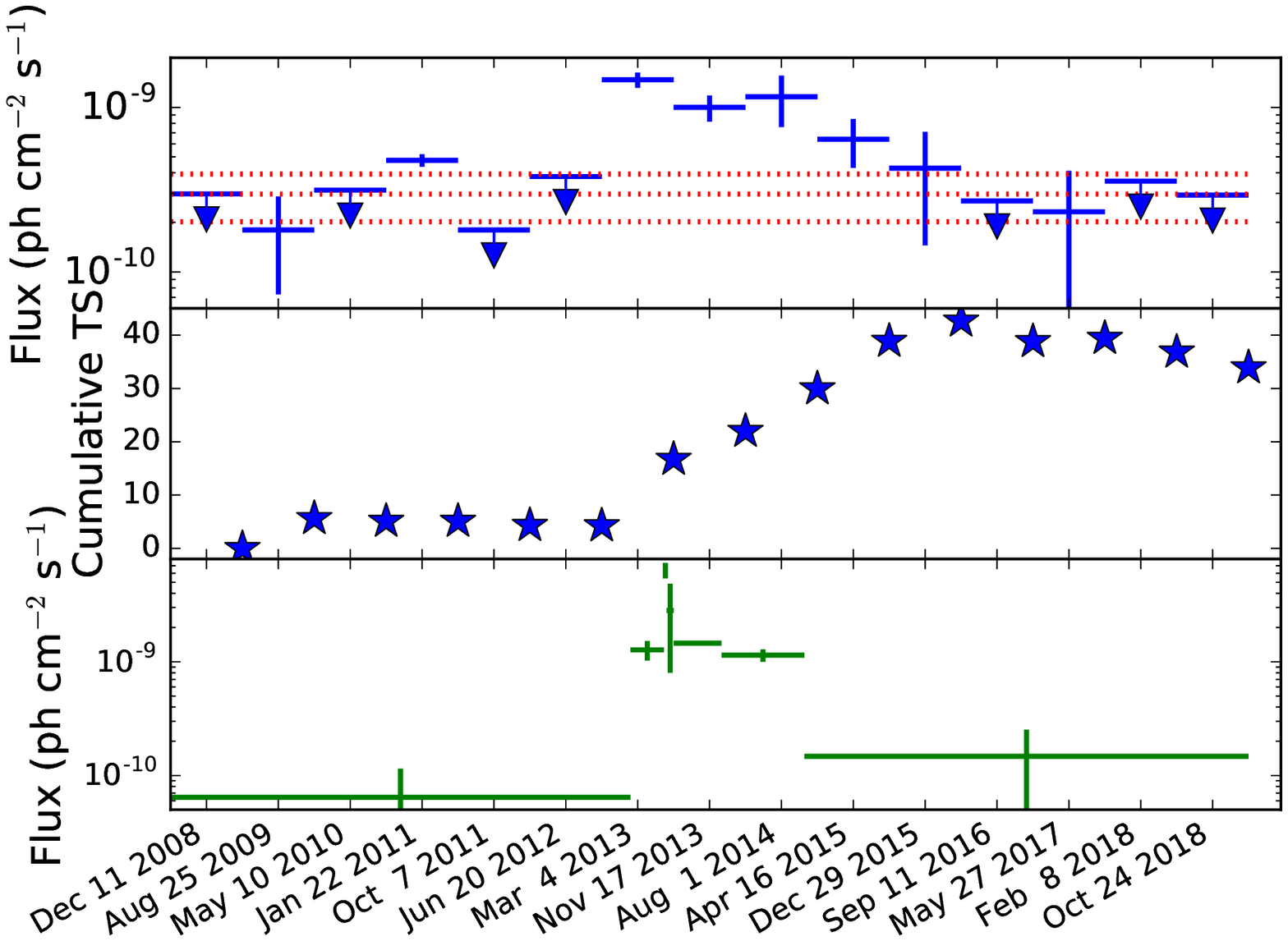}
\includegraphics[scale=0.45]{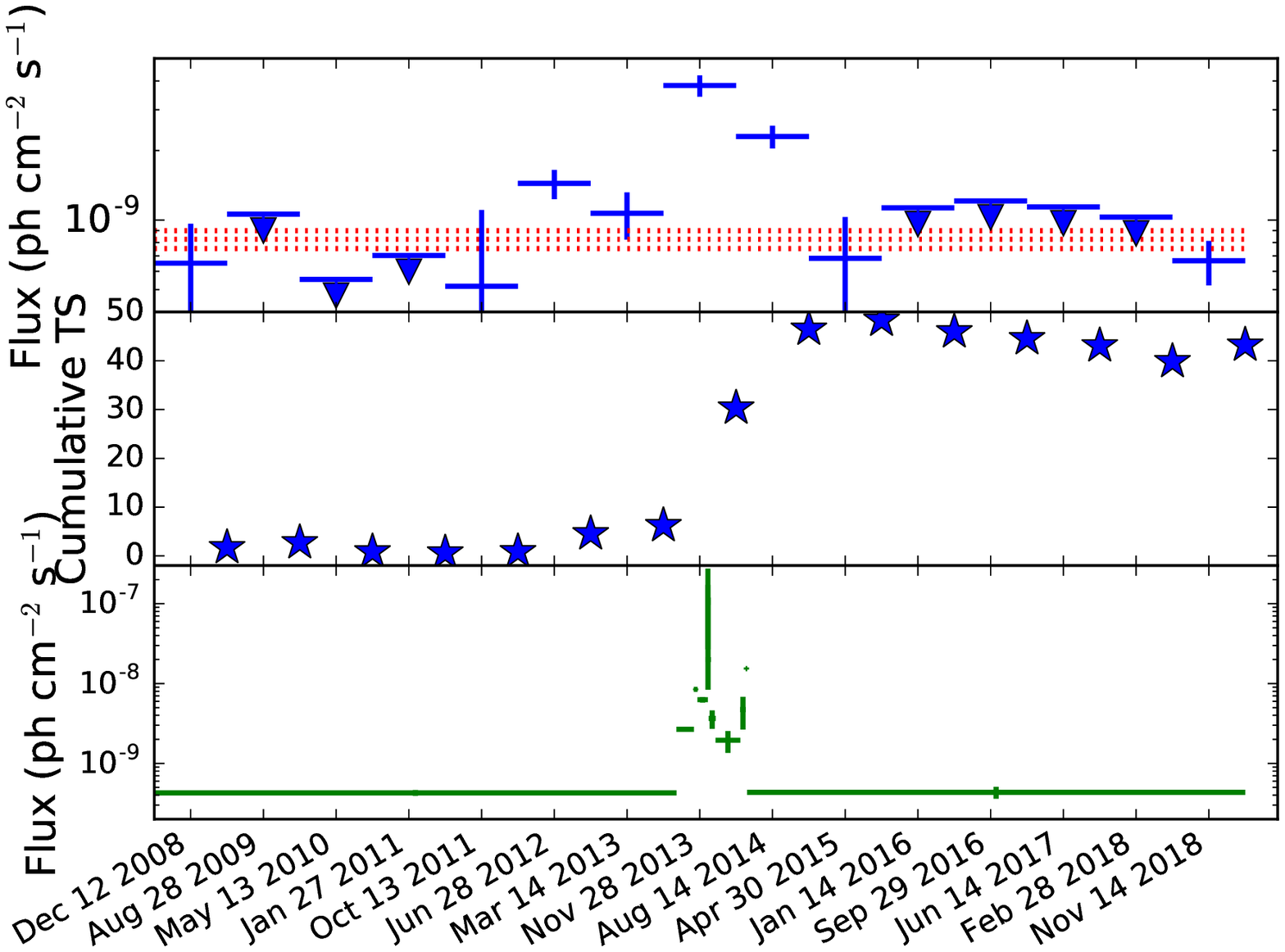}
\caption{Lightcurve and cumulative TS in $0.4-100$ GeV with fifteen time bins for NGC 3424 (left panel) and UGC 11041 (right panel). Top panels: the red dashed lines illustrate the maximum likelihood flux level for the $\sim 10.5$ year observations, the upper limits data points are at the 95\% confidence level. Middle panels: cumulative TS (blue star points) of the gamma-ray excess at the source position. Bottom panels: the green data points are generated with the adaptive-binning method based on constant $\rm TS = 9$.}
\label{figlc15}
\end{figure*}

\begin{figure*}
\centering
\includegraphics[scale=0.4]{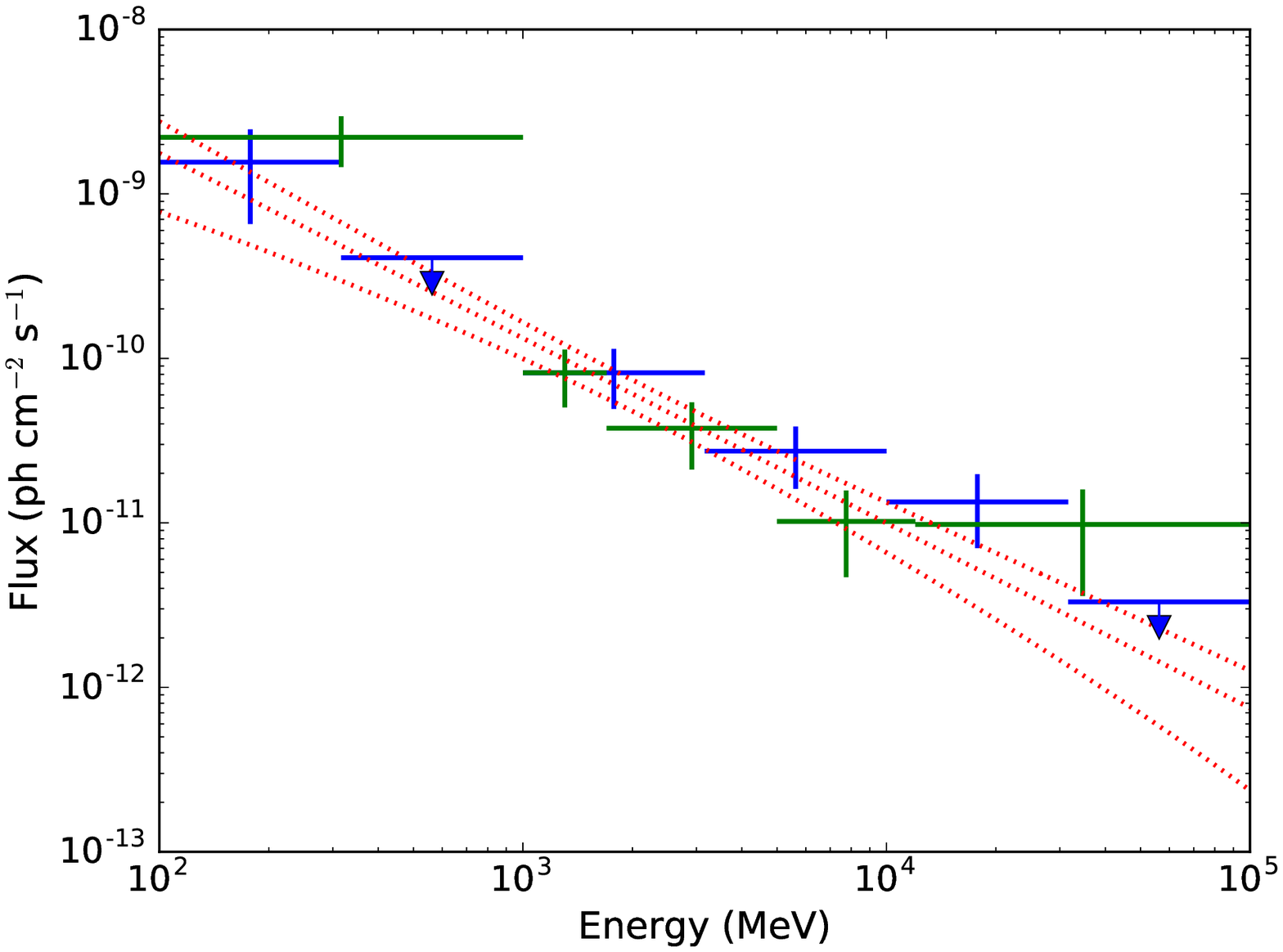}
\includegraphics[scale=0.4]{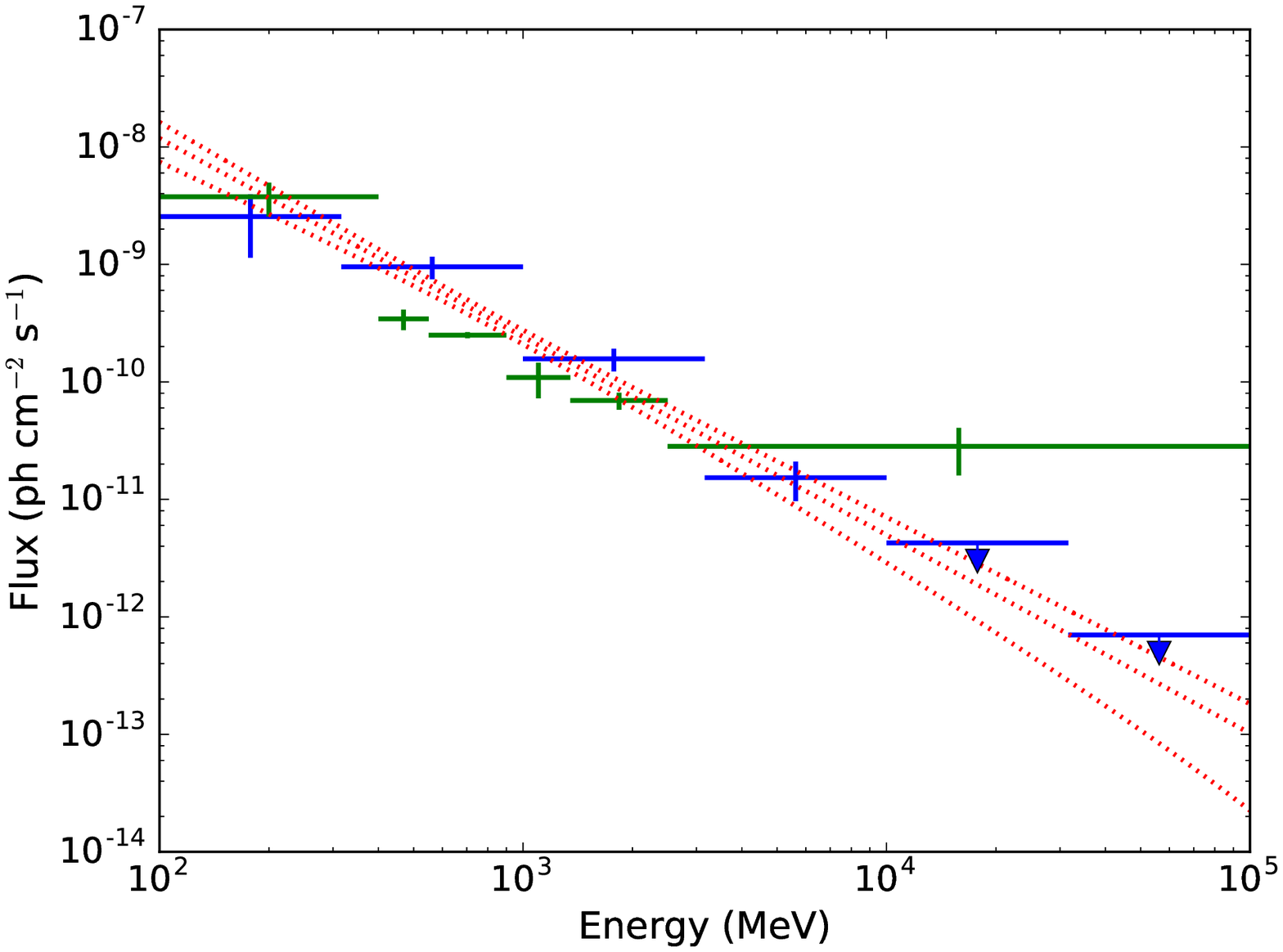}
\caption{Spectral energy distribution with six energy bins for NGC 3424 (left panel) and UGC 11041 (right panel). The red dashed lines illustrate the maximum likelihood flux level for the full energy band $0.1-100$ GeV. The upper limits data points are at the 95\% confidence level. The green data points are generated based on constant $\rm TS = 9$.}
\label{figsed}
\end{figure*}

We next checked a finer light curve for photon energies of 0.4-100 GeV with 15 time bins.
The bin number of $15$ is adopted for the balance of total detection significance and statistic.
The light curves and cumulative TS are shown in Figure \ref{figlc15}.
For NGC 3424, the $\chi^2$ goodness-of-fit test is inconsistent with a constant flux with a reduced $\chi^2$ of $2.14$ for data points with a detection threshold of $\rm TS > 1$, rejecting the hypothesis of constant flux.
Meanwile, the ratio between the maximum flux of 15 bins and full time-averaged flux (hereafter $\mathcal{R}$) reached $\sim 4.9$.
Since $\mathcal{R}$ is not affected by choice of various detection thresholds, $\mathcal{R} \simeq 4.9 $ indicates  another piece of solid evidence of variability.
The method most commonly used to quantify variability is a likelihood-based statistic.
Following the definition in 2FGL \citep{2012ApJS..199...31N}, the variability index from the likelihood analysis is constructed, with value in the null hypothesis, that the source flux is constant across the full time period, and the value under the alternate hypothesis where the flux in each bin is optimized:
\begin{equation}
TS_{\rm var} =  \sum_{i=1}^N 2\times ( Log(\mathcal{L}_{i}(F_i)) - Log(\mathcal{L}_{i}(F_{\rm mean}))),
\end{equation}
where $\mathcal{L}_i$ is the likelihood corresponding to bin $i$, $F_i$ is the best fit flux for bin $i$, and $F_{\rm mean}$ is the best fit flux for the full time, assuming a constant flux.
The statistic is expected to be distributed, in the null case, as $\chi^2_{N-1}$ ($TS_{\rm var}$).
For 15 bins, the critical value of $ TS_{\rm var} \geq 29.14$ is used to identify variable sources at a 99\% confidence level.
NGC 3424 should be flagged as variable, with $TS_{\rm var} = 45.3$.
Thus we claim that the gamma-ray emission of NGC 3424 is variable.
Similarly, for UGC 11041, the $\chi^2$ goodness-of-fit test was inconsistent with a constant flux with a reduced $\chi^2$ of $3.62$ for data points with a detection significance $\rm TS > 1$.
We derived $\mathcal{R}$ up to $\sim 4.6$, and the corresponding variability index $TS_{\rm var} = 54.1$, implying the presence of significant variability.
We also use the adaptive-binning method \citep{2012A&A...544A...6L} to generate the light curves based on constant $\rm TS = 9$, see Figure \ref{figlc15}. The gamma-ray flares of NGC 3424 and UGC 11041 are clearly detected.
One can see that the brightest flares of these two sources occur after the construction of the \textsl{Fermi}-LAT 3FGL catalog, so the two sources did not appear in \textsl{Fermi}-LAT 3FGL. 
The spectral energy distributions (SEDs) of the two sources are shown in Figure \ref{figsed}. No spectral curvature was detected.

We apply this approach to the gamma-ray emission of other starburst galaxies. We measured variability index $TS_{\rm var} = 22.0 $ and $\mathcal{R} \sim 1.6$ for NGC 1068, and $TS_{\rm var} = 31.8$ and $\mathcal{R} \sim 1.7$ for NGC 4945, meaning that NGC 1068 and NGC 4945 have no significant changes in flux over the  $10.5$ years of LAT observations.
We further took NGC 253, a typical starburst galaxy as an example to test our calculation.
We repeated the analysis and obtained variability index of $TS_{\rm var} = 22.8$ and $\mathcal{R} \sim 1.5$, supporting the lack of evidence for variability.
All these results are consistent with that in \citet{2012ApJ...755..164A}, implying that our approach is feasible.

\section{The outlier to the scale relations for $L_\gamma$}
There is a clear positive empirical relation between the gamma-ray luminosity and SFRs in local group galaxies and nearby star-forming/starburst galaxies \citep{2010A&A...523L...2A,2012ApJ...755..164A,2014ApJ...794...26T,2016ApJ...821L..20P,
2016ApJ...823L..17G,2017MNRAS.471.1737F,2019A&A...621A..70P,2019ApJ...874..173Z}.
The relation could be understood generally in the frame of connections among interstellar gas, star formation, and CRs (e.g., \citealp{2004ApJ...617..966T,2011ApJ...734..107L,2017ApJ...847L..13P,2018arXiv181210496C}).
Considering realistic galaxy properties, \citet{2019ApJ...874..173Z} find that the slope of the relation at low infrared luminosity deviates from that at high infrared luminosity end, due to increasingly lower efficiency for the production of gamma-ray emission.
From the point of statistics, the power law function (quasi-linear relationship in log-log space) is equally acceptable by the data \citep{2019ApJ...874..173Z}.
Thus, for simplicity, we still use the power law function in the following content for the same sample galaxies.
Now we check whether NGC 3424 and UGC 11041 obey the above relations.
Following the method in \citet{2019A&A...621A..70P}, we fit the relation of $L_{\gamma}-L_{\rm IR}$ and $L_{\gamma}-L_{\rm 1.4\ GHz}$ using the Markov Chain Monte Carlo code \textsl{emcee} \citep{2013PASP..125..306F}.
Compared with the previous results, there are some updates on the gamma-ray fluxes for LMC, M 31, NGC 1068 and NGC 4945 \citep{2019arXiv190210045T}.

\begin{figure*}
\centering
\includegraphics[scale=0.55]{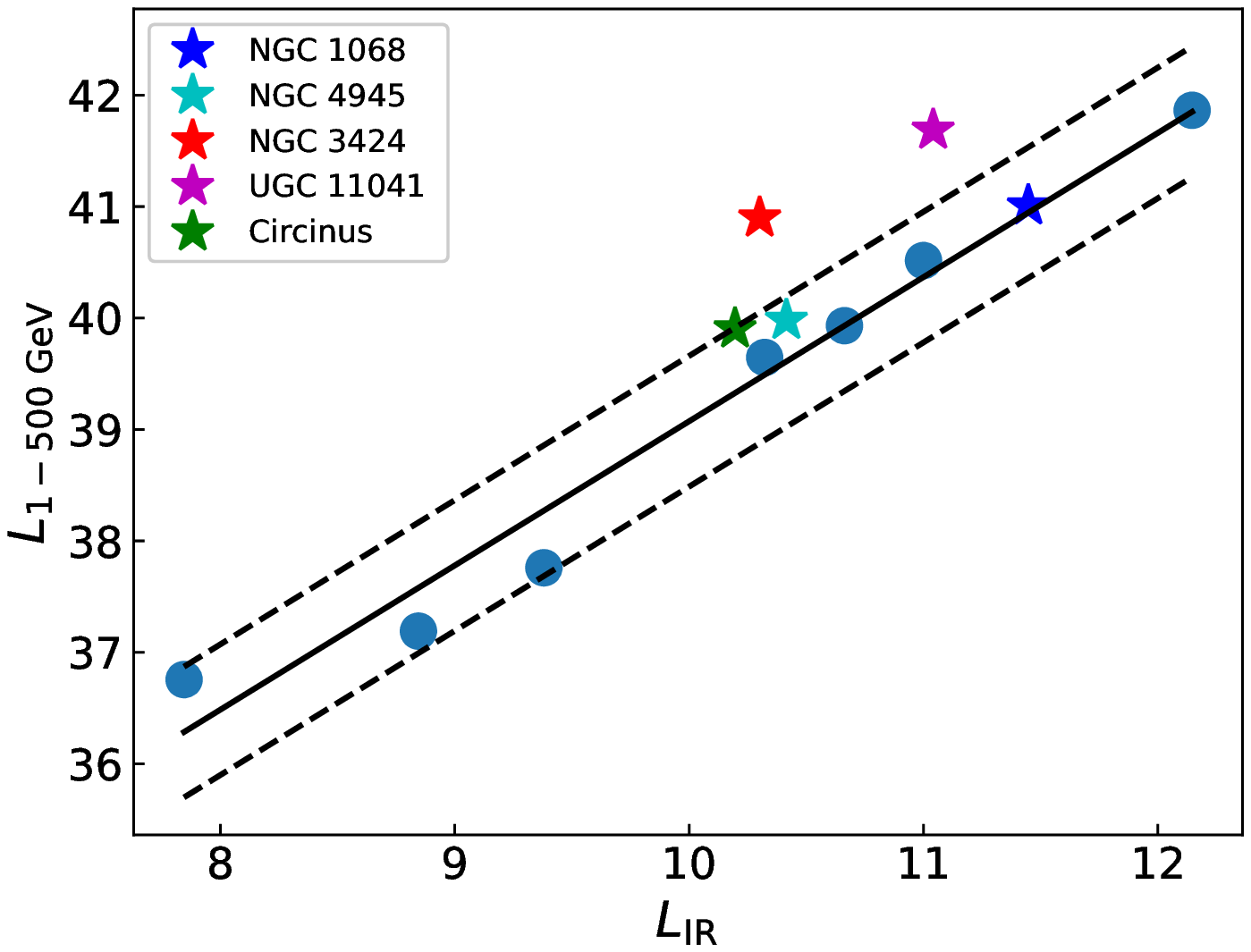}
\includegraphics[scale=0.55]{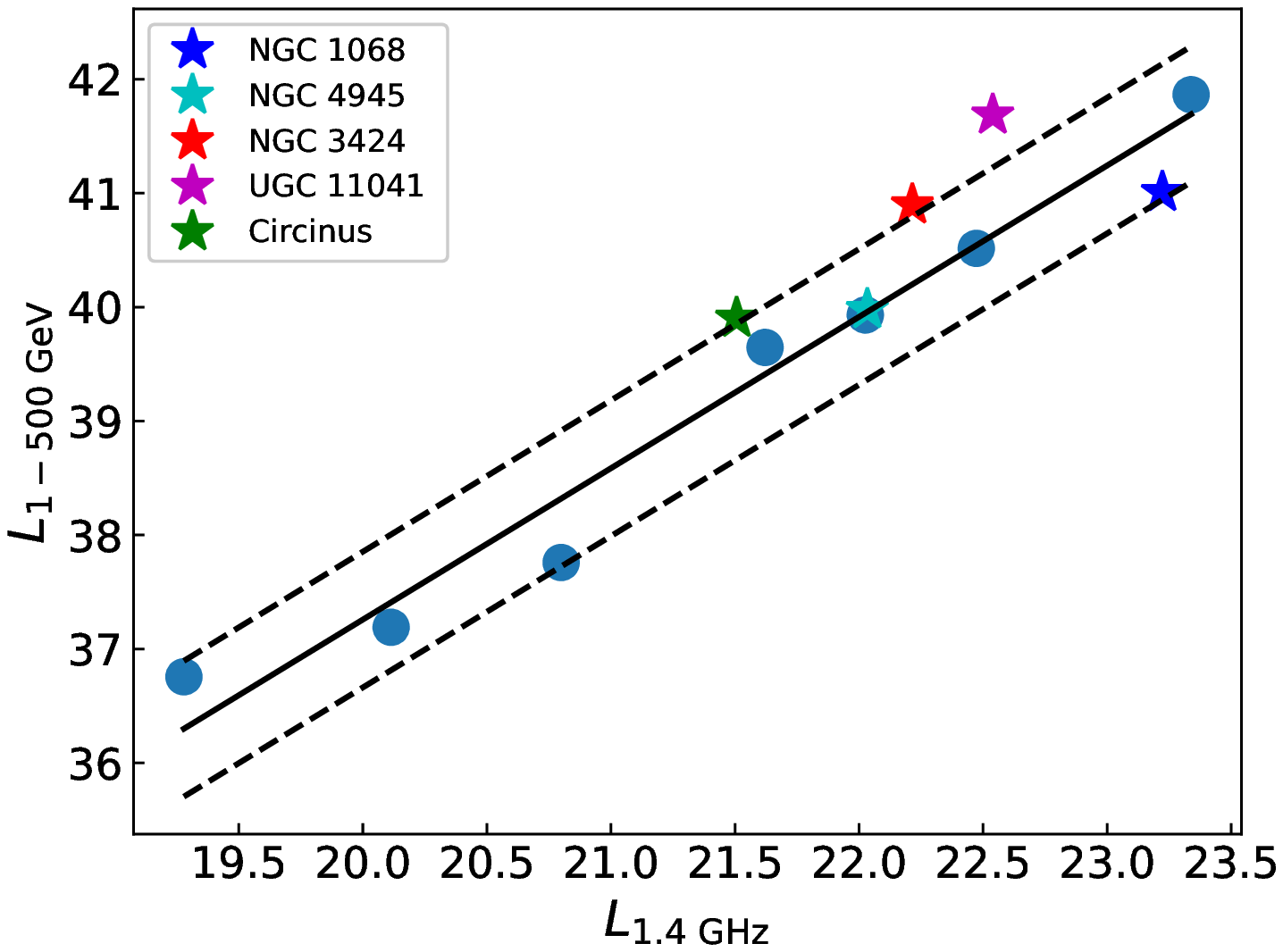}
\caption{ $L_{1-500 \ \rm GeV}-L_{\rm IR}$ relation (left panel) and $L_{1-500 \ \rm GeV}-L_{\rm 1.4 \ GHz}$ relation (right panel) for star-forming (and statburst) galaxies without detected AGN (shown in light blue circle data points).
The black solid line and dashed lines are the best-fit linear correlation and uncertainty at $1 \sigma$ confidence level, respectively.
The infrared luminosities in unit of $L_{\sun}$ are from \citet{2003AJ....126.1607S}. The radio continuum luminosities at 1.4 GHz in unit of $\rm W \ Hz^{-1}$ are from \citet{2001ApJ...554..803Y}, except for M82 \citep{1990ApJS...73..359C}, SMC \citep{1987A&A...178...62L}, LMC \citep{2007MNRAS.382..543H}, and M31 \citep{1975PASP...87...83D}.
The gamma-ray luminosities in unit of $\rm erg \ s^{-1}$ for starburst galaxies are taken from \citet{2019arXiv190210045T}.}
\label{lcLumIRRa}
\end{figure*}

The best-fit (logarithmic) relation between the updated gamma-ray and IR (and monochromatic radio continuum) luminosities of a sample of pure star-forming galaxies, excluding galaxies with AGN (i.e., NGC 1068, NGC 4945, NGC 3424 and UGC 11041), are ${\rm log} L_{1-500 \ \rm GeV} = (26.46^{+1.32}_{-1.30}) + (1.29^{+0.13}_{-0.13}) {\rm log} L_{\rm IR}$ (${\rm log} L_{1-500 \ \rm GeV} = (10.69^{+2.85}_{-2.87}) + (1.33^{+0.13}_{-0.13}) {\rm log} L_{1.4 \ \rm GHz}$).
Both the Pearson correlation coefficients $r > 0.9$ and chance probabilities $p < 10^{-4}$ imply strong correlations.
The relation of $L_{1-500 \ \rm GeV}-L_{\rm IR}$ and $L_{1-500 \ \rm GeV}-L_{\rm 1.4\ GHz}$ are shown in Figure \ref{lcLumIRRa}.
The two new sources, NGC 3424 and UGC 11041 appear as  remarkable outliers of the $L_{1-500 \ \rm GeV}-L_{\rm IR}$ correlation.
For $L_{1-500 \ \rm GeV}-L_{\rm 1.4\ GHz}$ correlation,  UGC 11041 appears as an outlier, but NGC 3424 as marginal outlier located at the boundary line of $1 \sigma$ dispersion region.
In contrast, two other starburst galaxies with radio-quiet AGNs (i.e., NGC 1068 and NGC 4945)  obey the $L_{1-500 \ \rm GeV}-L_{\rm IR}$ ($L_{1-500 \ \rm GeV}-L_{\rm 1.4\ GHz}$) correlations. This may indicate that the gamma-ray emission NGC 1068 and NGC 4945 is mainly contributed by CR interactions \citep{2010A&A...524A..72L,2012ApJ...755..164A,2016A&A...596A..68L}.
The properties of starburst galaxies with detected AGNs are summarized in Table \ref{tablesfgagn}.

\begin{table*}[!htp]
\centering
\caption{Summary of star-forming (and starburst) galaxies with detected AGN}
\begin{tabular}{cccccc}
\hline
\hline
Name &  NGC 1068 & NGC 4945 & NGC 3424 & UGC 11041 & Circinus\\
\hline
Dis. & 16.7 & 3.7 &  26.2 & 69.9 & 4.2\\
\hline
$L_{\rm IR}$ &11.45 & 10.41 & 10.30 & 11.04 & 10.20\\
\hline
$L_{\rm 1.4\ GHz}$&  167 & 10.8 & 16.4 & 34.6 & 3.21\\
 & $\times 10^{21}$ & $\times 10^{21}$ & $\times 10^{21}$  & $\times 10^{21}$ & $\times 10^{21}$ \\
\hline
$L_{1-500\ \rm GeV}$& $1.02\pm 0.14$ & $9.57\pm 0.97$ & $7.93\pm 2.47$ & $4.87\pm 1.14$ & $8.02\pm 1.27$\\
  & $\times 10^{41}$ & $\times 10^{39}$ & $\times 10^{40}$  & $\times 10^{41}$ & $\times 10^{39}$\\
\hline
$TS_{\rm var}$ & 22.0 & 31.8 & 45.3 & 54.1 & 16.1\\
\hline
\hline
\end{tabular}

\textbf{Notes.} The second row is distance in unit of Mpc,  NGC 1068 and NGC 4945  \citep{2004ApJ...606..271G}, NGC 3424 and UGC 11041 \citep{2003AJ....126.1607S}, and Circinus galaxy \citep{2009AJ....138..323T}. The third row is $L_{\rm IR} = L_{8-1000 \ \micron }$ with unit of $L_{\sun}$ in logarithmic space, NGC 1068 and NGC 4945  \citep{2004ApJ...606..271G}, NGC 3424 and UGC 11041 \citep{2003AJ....126.1607S}, Circinus galaxy\citep{1996ARA&A..34..749S}. The radio continuum luminosity at 1.4 GHz $L_{\rm 1.4\ GHz}$ with unit of $\rm W \ Hz^{-1}$ is from \citet{2001ApJ...554..803Y} for NGC 1068 and Circinus galaxy, \citet{1990PKS...C......0W} for NGC 4945, \citet{1992ApJS...79..331W} for NGC 3424, and \citet{1998AJ....115.1693C} for UGC 11041. The time-averaged gamma-ray luminosity with unit of $\rm erg \ s^{-1}$ are taken from \citet{2019arXiv190210045T}, except for NGC 3424 and UGC 11041. The $TS_{\rm var}$ is calculated using equation (1), the critical value of $ TS_{\rm var} \geq 29.14$ is used to identify variable source at a 99\% confidence level for 15 time bins.
\label{tablesfgagn}
\end{table*}

\section{Discussions}
The gamma-ray properties of NGC 3424 and UGC 11041, including significant variability and bright gamma-ray luminosity, imply that the radiation processes are different from other starburst galaxies, whose gamma-ray emission is mainly from CRs diffuse process.
This suggests that their high energy gamma-ray emissions require an extra contribution.
Therefore we suggest NGC 3424 and UGC 11041 may harbor obscured AGNs, and attribute the gamma-ray emission associated with NGC 3424 and UGC 11041 to originate from AGN activity.
For NGC 3424, this is supported by the fact that NGC 3424 is classified as a AGN based on the [NII]/$\rm H_{\alpha}$ ratio combined with the strength of the $\rm H_{\alpha}$ line \citep{2013A&A...558A..68G}.
Interestingly, NGC 3424 appears to be in the early stage of  merger with NGC 3430 \citep{2013ApJ...768...90L}, which is capable of triggering not only AGN activity, but also star formation process.

The major part of gamma-ray sources in 3FGL are radio-loud AGNs \citep{2015ApJ...810...14A}, which are observed to host  powerful relativistic jets.
However, NGC 3424 has a ratio of radio (5 GHz) to optical (B-band) flux $F_5/F_B < 10$, meaning that it is a radio-quiet AGN \citep{1989AJ.....98.1195K}\footnote{\url{http://ned.ipac.caltech.edu/}}.
According to the [3.4]-[4.6]-[12] $\mu$m color-color diagram observed by WISE \citep{2010AJ....140.1868W}, NGC 3424 is located in the starburst region, implying that the IR emission is not from the non-thermal process.
We also note that, different from $L_{1-500 \ \rm GeV}-L_{\rm IR}$ relation, NGC 3424 is marginally consistent with the $L_{1-500 \ \rm GeV}-L_{\rm 1.4\ GHz}$ correlation (see Figure \ref{lcLumIRRa}), which is not well-understood and may require further investigation.
As for UGC 11041, besides the WISE data demonstrating that the IR emission is probably from thermal emission as well, there is not much information available, so further investigation is necessary.

The situation of NGC 3424 is somewhat similar to  Circinus galaxy, a system with coexistence of a typical obscured Seyfert-type active nucleus and a starburst.
As one of the closest known active galaxies to the Milky Way, the Circinus galaxy is located behind the intense foreground of the Milky Way disk, $(l, b) = (311\arcdeg .3, -3\arcdeg .8)$, resulting in much uncertainty and ambiguity with \textsl{Fermi}-LAT data analysis.
With respect to the results in \citet{2013ApJ...779..131H}, the updated gamma-ray flux has been reduced roughly by half \citep{2019arXiv190210045T}. 
Thus, its $L_{\gamma}$ value moved closer to this boundary line of 1$\sigma $ dispersion region of the correlations, see Figures \ref{lcLumIRRa}.
Similarly, we measured the $TS_{\rm var} = 16.1$ and $\mathcal{R} \sim 1.7$ for 15 bins light curve.
These characteristics favor that the gamma-ray emission originates from CR interactions with ISM, see \citet{2019arXiv190504723G} for further information.
Interestingly, a possible companion to the Circinus galaxy has been found \citep{2016AJ....151...52S}, which could explain the star formation activity in Circinus induced by a strong tidal perturbation in this binary galaxy system.

In addition to NGC 3424 and Circinus, starburst galaxies in our sample once or now are probably undergoing major or minor mergers.
Major merger, minor merger, or transient encounter may trigger circumnuclear starburst or(and) AGN, although the companion in this so-called starburst-seyfert system is not necessarily discernible \citep{2004ApJ...605..144M}.
Arp 220, for example, dense gas in the circumnuclear region induced by strong interaction during major merger of two disc galaxies (e.g., \citealp{1990ApJ...354L...5G}), leads to intense star formation activity.
Although a low-luminosity AGN could be present \citep{2015ApJ...814...56T,2017MNRAS.469L..89Y,2018MNRAS.474.4073W,2019MNRAS.484.3665Y}, star formation process contribute a dominant fraction of the gamma-ray radiation in Arp 220 \citep{2016ApJ...821L..20P}.
These systems where gamma-ray emission is dominated by processes associated with the circumnuclear star formation rather than with the AGN should be recognized as starburst-dominant Seyfert galaxies in the framework of \citet{2004ApJ...605..144M}.  We could further speculate that the fraction of gamma-ray emission powered by AGN will increase with the consumption of gas in starburst galaxies like Arp 220.
NGC 3424 with significant indication of variability is clearly different from these starburst galaxies.
The evidence for AGN in NGC 3424 \citep{2013A&A...558A..68G} is consistent with the elevated gas supply and nuclear activity expected during the interaction between NGC 3424 and NGC 3430, despite the early stage of merger. The star formation activity in NGC 3424 may become more intense in later stages of the merger, and the system probably evolves to a starburst-dominant Seyfert galaxy if the star formation rate is high enough.

\section{Conclusions}
We performed a detailed analysis of \textsl{Fermi}-LAT data  around the regions of two ordinary IRAS sources, NGC 3424 and UGC 11041, which have recently been reported to be associated with 4FGL  sources \citep{2019arXiv190210045T}.
We paid special attention to the flux variability. Significant  temporal variability is evident in the gamma-ray emission of NGC 3424 and UGC 11041.
We calculated the gamma-ray luminosities and investigated whether NGC 3424 and UGC 11041 obey the empirical $L_{\gamma}-L_{\rm IR}$ and $L_{\gamma}-L_{\rm 1.4\ GHz}$ relations of starburst galaxies.
We find that NGC 3424 and UGC 11041 appear as outliers of $L_{\gamma}-L_{\rm IR}$ and $L_{\gamma}-L_{\rm 1.4\ GHz}$ relations (though NGC 3424 locates at the boundary of $L_{\gamma}-L_{\rm 1.4\ GHz}$ correlation).
Their gamma-ray luminosities are about  an order of magnitude larger than  what are expected from CR interactions.
These results suggest that NGC 3424 and UGC 11041 may host AGNs, and the gamma-ray emission of these two starburst galaxies may arise from the AGN contribution, not from star-formation driven CR process.
Finally, for the gamma-ray excess around UGC 11041, we note that the association with a flat-spectrum radio source MG2 J175448+3442 could not be excluded.

\vspace{1 cm}
X.Y. Wang is supported by the National Key R \& D program of China under the grant 2018YFA0404203 and the NSFC  grants 11625312 and 11851304.
J.-F. Wang is supported by the National Key R\&D Program of China (2016YFA0400702) and the NSFC grants U1831205, 11522323.
F.K. Peng acknowledges support from the Doctoral Starting up Foundation of Guizhou Normal University 2017 (GZNUD[2017] 33) and the Technology Department of Guizhou province Fund under grant No. [2019]1220.
Q.J. Zhi acknowledges support from the science and technology innovation talent team (grant (2015)0415), the High Level Creative Talents (grant (2016)-4008) and Innovation Team Foundation of the Education Department (grant [2014]35) of Guizhou Province.
This research has made use of the NASA/ IPAC Infrared Science Archive, which is operated by the Jet Propulsion Laboratory, California Institute of Technology, under contract with the National Aeronautics and Space Administration.
The authors are grateful to the anonymous referees for the useful and detailed comments.
F.K. Peng thanks M.-Y. Xiao, D.-H. Yan, and Q.-C. Liu for useful discussions.

\facility{\textsl{Fermi}(LAT)}

\section*{ORCID iDs}
Fang-Kun Peng: \url{https://orcid.org/0000-0001-7171-5132}

\end{CJK*}
\end{document}